# Generalizing and transporting causal inferences from randomized trials in the presence of trial engagement effects

Lawson Ung[*1,2], Tyler J. VanderWeele[1-3], and Issa J. Dahabreh[1-3]

[1]CAUSALab, Harvard T.H. Chan School of Public Health, Boston, MA

[2]Department of Epidemiology, Harvard T.H. Chan School of Public Health, Boston, MA

[3]Department of Biostatistics, Harvard T.H. Chan School of Public Health, Boston, MA

Friday 19th July, 2024

**Address for correspondence:** Dr. Lawson Ung, Department of Epidemiology, Harvard T.H. Chan School of Public Health, Boston, MA 02115; email: lawson_ung@hsph.harvard.edu; phone: +1 (617) 495-1000.

---

[*]Address for correspondence: Dr. Lawson Ung, Department of Epidemiology, Harvard T.H. Chan School of Public Health, Boston, MA 02115; email: lawson_ung@hsph.harvard.edu; phone: +1 (617) 495-1000.

**Abstract**

Trial engagement effects are effects of trial participation on the outcome that are not mediated by treatment assignment. Most work on extending (generalizing or transporting) causal inferences from a randomized trial to a target population has, explicitly or implicitly, assumed that trial engagement effects are absent, allowing evidence about the effects of the treatments examined in trials to be applied to non-experimental settings. Here, we define novel causal estimands and present identification results for generalizability and transportability analyses in the presence of trial engagement effects. Our approach allows for trial engagement effects under assumptions of no causal interaction between trial participation and treatment assignment on the absolute or relative scales. We show that under these assumptions, even in the presence of trial engagement effects, the trial data can be combined with covariate data from the target population to identify average treatment effects in the context of usual care as implemented in the target population (i.e., outside the experimental setting). The identifying observed data functionals under these no-interaction assumptions are the same as those obtained under the stronger identifiability conditions that have been invoked in prior work. Therefore, our results suggest a new interpretation for previously proposed generalizability and transportability estimators; this interpretation may be useful in analyses under causal structures where background knowledge suggests that trial engagement effects are present but interactions between trial participation and treatment are negligible.

# INTRODUCTION

In the growing literature on extending (generalizing and transporting) causal inferences from a randomized trial to a target population of substantive interest [1–7], it is often, explicitly or implicitly, assumed that treatment effects can be studied without concern for trial engagement effects. Trial engagement effects are effects on the outcome from the invitation to participate in a trial or from trial participation itself that operate through paths that do not involve treatment assignment [8,9]. Structural models used in generalizability and transportability analyses [8,10] usually assume the absence of trial engagement effects by imposing an exclusion restriction from trial participation to the outcome of interest [8], thereby rendering causal estimands context-independent after accounting for differences in baseline covariates between the trial and the target population [5].

In many settings, however, the assumption that trial engagement effects are absent may not be plausible. For example, it has long been understood that trial participation itself may affect outcomes through paths not mediated by randomized treatment assignment, including paths that capture behavioral modifications (e.g., Hawthorne effects [11,12]), ancillary medical care [13–15], and mechanisms that promote treatment adherence (e.g., through regular contact with the research team [9]) among trial participants. In such settings, the effect of outcome-relevant trial procedures may be to such an extent that trial participation could be thought of as an intervention separate to treatment assignment. Ostensibly, without treatment and outcome data from the target population, the presence of engagement effects would render the effect of treatment unidentifiable among nonrandomized individuals not subject to the processes and procedures adopted in the trial.

Here, to address this problem, we define novel causal estimands for generalizability and transportability analyses that allow for the presence of trial engagement effects by considering joint intervention on trial participation and treatment assignment. We focus specifically on the average treatment effect in the context of usual care as implemented in the target population (i.e., outside the experimental setting), and present novel identification results that assume no causal interaction between trial participation and treatment assignment on the absolute or relative scales [16–19]. Interestingly, the identifying observed data functionals under these no-interaction conditions are the same as those obtained under stronger identifiability conditions that have been invoked in past work [4,5,20], suggesting a new interpretation for previously proposed estimators in applications where trial engagement effects are present but there is no interaction between trial participation and treatment assignment.

# 1 STUDY DESIGN, SAMPLING, AND DATA STRUCTURES

## 1.1 Study design and sampling

**The trial and the target population:** Consider a randomized trial (henceforth referred to as the "index" trial) evaluating two or more treatments. The randomized trial comprises individuals who are invited to participate, agree to enroll, and are subject to and engage with the activities (e.g., ancillary procedures and follow-up schedules) that accompany trial participation. The target population is a population of substantive interest to which the results of the randomized trial will be applied. Depending on the investigators' scientific goals, the target population may be defined by the eligibility criteria applied in the trial, as in generalizability analyses; alternatively, the target population may be defined by broadening the trial eligibility criteria, as in transportability analyses [21, 22]. Nonrandomized individuals in the target population do not participate in the trial, either because they were not invited or were invited but chose to decline, and are thus subject to usual or routine care practices that occur outside trial settings.

To focus on key issues concerning trial engagement effects, we make some simplifying assumptions. We do not distinguish between the invitation to participate in the trial and trial participation itself, because this has been addressed in prior work [8]. We also assume no losses to follow-up and complete adherence to treatment (such that the treatment assigned is the treatment received) [9]. Complexities related to loss to follow-up and non-adherence can be addressed using previously described approaches for censoring and time-varying treatments [23, 24], but we do not pursue these extensions here.

**Sampling:** Under the sampling scheme of *nested* trial designs, trial participants may be enrolled within a cohort (e.g., a healthcare system or registry [25]) that is regarded as a simple random sample of the underlying target population of interest [26], allowing the identification of average causal effects in the entire target population as well as the subset of nonrandomized individuals [5, 6]. In contrast, under the sampling scheme of *non-nested* trial designs, data from the trial and nonrandomized individuals are sampled separately from their underlying populations (with unknown sampling fractions), precluding the identification of causal effects in the entire target population, but still allowing the identification of effects in the nonrandomized subset of the target population [6]. In the main text, we present results for nested trial designs, specifically those that pertain to the entire target population comprising trial participants and nonparticipants. In the Supplement, we provide results that pertain to the subset of nonparticipants in nested trial designs; these results, with slight modifications, can also be applied to non-nested trial designs [5].

## 1.2 Notation

Let $X$ denote baseline (pre-treatment) covariates; $S$ a binary indicator for trial participation status (1 for participants in the index trial who are subject to trial scale-up activities that follow study enrollment, and 0 for non-trial participants who are subject to usual standards of care that occur outside trial settings); $A$ a binary indicator for assigned treatment (1 for the experimental treatment; 0 for the comparator treatment); a non-continuous, non-failure time outcome $Y$ measured at the end of the study; and unmeasured covariates $U$. We only consider binary treatments, though our results can be extended to treatments with multiple levels.

## 1.3 Data structure

As in most generalizability and transportability analyses, we require data on covariates $X$, treatment $A$, and outcome $Y$ from trial participants ($S = 1$), but we only require data on baseline covariates $X$ from non-participants ($S = 0$). In this setting, the identification results will apply even if treatments adopted in the index trial are different to those adopted in the target population. A schematic depiction of the composite dataset structure is as follows.

| | | | |
|---|---|---|---|
| $X$ | $S = 1$ | $A = 1$ | $Y$ |
| | | $A = 0$ | |
| | $S = 0$ | Not required | Not required |

## 2 CAUSAL MODEL AND ESTIMANDS

### 2.1 Counterfactual outcomes

To define causal estimands of interest and discuss identifiability conditions, we use counterfactual (potential) outcomes [27–30] under a finest fully randomized causally interpretable structured tree graph (FFRCISTG) causal model [29, 31, 32]. We examine interventions on trial participation: $S$ is set to $s = 1$ to denote the scale-up of trial-related activities that follow trial enrollment, and set to $s = 0$ to denote usual or routine care practices that occur outside trial settings. We also examine intervention on treatment $A$ set to $A = a$, taking values of either $a = 1$ to denote an experimental treatment, or $a = 0$ to denote a "control" treatment.

Under our causal model, counterfactuals induced by intervening on any model variable are considered well-defined, and consistency holds for all of these interventions. Specifically, we consider the following variables well-defined: $A^{s=0}$ as the counterfactual treatment assignment under intervention to set trial participation $S$ to $s = 0$; $A^{s=1}$ as the counterfactual treatment assignment under intervention to set trial participation $S$ to $s = 1$; $Y^a$ as the counterfactual outcome under intervention to set treatment $A$ to $a$; $Y^{s=1,a}$ as the counterfactual outcome under joint intervention to set trial participation $S$ to $s = 1$, and treatment $A$ to $A = a$; and finally $Y^{s=0,a}$ as the counterfactual outcome under joint intervention to set trial participation $S$ to $s = 0$, and treatment $A$ to $A = a$.

### 2.2 Causal estimands

In a nested trial design [6], investigators might be interested in identifying (and estimating) average treatment effects under different assigned treatments in the target population comprising both randomized and nonrandomized individuals, as captured in $\mathrm{E}[Y^a - Y^{a'}]$. This causal quantity is "context-independent" in the sense that it does not involve intervening on trial participation and therefore is most interesting when trial engagement effects are absent.

If trial engagement effects are present, other causal estimands that also involve intervention on trial participation – where one intervenes on outcome-relevant trial activities other than treatment – might be of interest. For example, investigators might be interested in the treatment effect for the randomized treatment under some pragmatic scenario where individuals are subject to some usual standard of care and real-world clinical practices adopted outside trial settings (i.e., where one jointly intervenes to set trial participation $S$ to $s = 0$ and treatment $A$ to $a$). In this case, the relevant estimand is the average treatment effect in the context of usual care as used in the target population, $\mathrm{E}[Y^{s=0,a} - Y^{s=0,a'}]$. Here, "usual care" should be interpreted as the outcome-relevant aspects of care as used in the target population, outside the experimental context.

Alternatively, investigators might be interested in the treatment effect under scenarios where all outcome-relevant trial activities, in addition to treatment, are scaled up in the target population (i.e., where one jointly intervenes to set trial participation $S$ to $s = 1$ and treatment $A$ to $a$). Here, the relevant estimand is the average treatment effect in the context of trial protocol-based care, $\mathrm{E}[Y^{s=1,a} - Y^{s=1,a'}]$. Of note, in this paper we focus on average treatment effects, rather than their component potential outcome means; the identification results we obtain, based on assumptions of no causal interaction between trial participation and treatment on the absolute scale, suffice to identify the former but not the latter.

## 2.3 Graphical representation of the causal structure

The causal directed acyclic graph (DAG) presented in Figure 1 represents the causal structure of the target population, comprising both trial and non-trial participants. The DAG has features that are typical of previous work on generalizability and transportability methods. First, there is no unmeasured common cause of trial participation $S$ and the outcome $Y$ (e.g., no $U \to S$ arrow); informally, there are no unmeasured covariates that predict whether an individual would choose to participate in the trial that also affect the outcome. Second, trial participation $S$ has an effect on treatment $A$, depicted by the $S \to A$ arrow to indicate the different treatment assignment mechanism among the subsets of trial participants and nonparticipants. Third, unmeasured variables $U$ are common causes of treatment $A$ and the outcome $Y$, reflecting the absence of treatment randomization in the nonrandomized subset of the target population. In contrast to prior work [4, 5, 8, 9], however, we allow for trial engagement effects on the outcome, represented by the $S \to Y$ arrow. Informally, these trial engagement effects can include the effects of trial participation itself, or outcome-relevant trial activities for which trial participation can serve as a proxy.

## 3 CORE IDENTIFIABILITY CONDITIONS

We begin by formalizing core identifiability conditions encoded in the causal model represented in Figures 1, 3, and 4 (conditions A1, A2, and A3 listed below), as well as positivity conditions that have been invoked in prior work on generalizability and transportability analyses. This set of assumptions is used repeatedly in the following sections.

A1. *Consistency of potential outcomes with respect to interventions on trial participation $S$ and treatment $A$:* If $S_i = s$ and $A_i = a$, then $Y_i^{s,a} = Y_i$.

Condition A1 states that the potential outcome for the $i^{th}$ individual under intervention to set trial participation to $S = s$ and treatment to $A = a$ is the same as the observed outcome for that individual under the same observed trial participation and treatment levels. Unlike in previous transportability work [4, 5, 9, 33], we do *not* assume that trial engagement effects are absent; that is to say, we do *not* assume $Y_i^{s,a} = Y_i^a$ for any $s$ and any $a$.

A2. *Conditional exchangeability (transportability) over trial participation $S$:* Under the proposed causal model, $Y^{s=1,a} \perp\!\!\!\perp S | X$.

Condition A2 states that, for the joint intervention setting trial participation $S$ to $s = 1$ and treatment $A$ to $a$, the distribution of potential outcomes is the same for trial and non-trial participants, conditional on covariates $X$. This condition implies a weaker mean exchangeability condition: for every covariate pattern $x$ with positive density $f(x) \neq 0$ in the target population, $\mathrm{E}\left[Y^{s=1,a}|X = x\right] = \mathrm{E}\left[Y^{s=1,a}|X = x, S = 1\right]$.

We note in passing that under our causal model, an exchangeability condition stronger than condition A2 holds: $Y^{s,a} \perp\!\!\!\perp S | X$ for $s \in \{0, 1\}$ (see Figures 3 and 4). In the main text, however, we only use condition A2.

A3. *Conditional exchangeability over treatment $A$ in the trial:* Under the proposed causal model, and under consistency condition A1, $Y^{s=1,a} \perp\!\!\!\perp A | (X, S = 1)$.

Condition A3 states that, for the joint intervention setting trial participation $S$ to $s = 1$ and treatment $A$ to $a$, the distribution of potential outcomes is the same between observed treatment levels, conditional on covariates $X$. This condition is expected to hold if the index trial has been marginally randomized. This condition also implies a weaker mean exchangeability condition, that is for every covariate pattern $x$ with positive density $f(x, S = 1) \neq 0$ in the trial population, $\mathrm{E}[Y^{s=1,a}|X = x, S = 1] = \mathrm{E}[Y^{s=1,a}|X = x, S = 1, A = a]$.

A4. *Positivity of treatment assignment $A$ in the trial:* For every covariate pattern $x$ with positive density $f(x, S = 1) \neq 0$ in the trial population and for each $a \in \mathcal{A}_{S=1}$, $\Pr[A = a|X = x, S = 1] > 0$.

Condition A4 states that in the index trial, there is positive probability of being assigned to each level of treatment $A$, conditional on covariates $X$ required for conditional exchangeability. This condition is expected to hold by design in a randomized trial, whether marginally randomized or randomized conditional on covariates $X$.

A5. *Positivity of trial participation:* For every covariate pattern $x$ with positive density $f(x) \neq 0$ in the target population, there is a positive probability of trial participation, that is $\Pr[S = 1|X = x] > 0$.

Condition A5 states that the covariate patterns $x$ needed for conditional exchangeability over trial participation $S$ must have common support (i.e., sufficient overlap) between the populations underlying the trial and the nonrandomized target population.

## 4 IDENTIFICATION CHALLENGES WHEN TRIAL ENGAGEMENT EFFECTS ARE PRESENT

We now discuss some challenges in identifying relevant causal estimands in the presence of trial engagement effects, unless one is willing to make additional assumptions. To do so, we study a series of single world intervention graphs (SWIGs) [32, 34]. Specifically, starting from the same directed acyclic graph (Figure 1) we draw SWIGs under different interventions: intervention to set treatment $A$ to $a$ (Figure 2); joint intervention to set trial participation $S$ to $s = 0$ and treatment $A$ is set to $a$ (Figure 3); and joint intervention to set trial participation $S$ to $s = 1$ and treatment $A$ to $a$ (Figure 4). We note in passing that the graphs we consider are more precisely termed single world intervention *templates* because the node representing $a$ can take multiple values [32, 34]; this slight abuse of terminology is common in the literature and does not cause much confusion.

Because the goal of generalizability and transportability analyses is usually to identify and estimate average treatment effects in a target population (or its subsets), without regard to a specific (experimental or non-experimental) setting, one might simply propose intervening to set treatment $A$ to $a$ (Figure 2). However, the independencies $Y^a \perp\!\!\!\perp A|X$ and $Y^a \perp\!\!\!\perp A|(X,S)$ are not expected to hold under this graph because of the $A \leftarrow U \rightarrow Y^a$ path, reflecting the presence of unmeasured confounding in the nonrandomized subset of the target population. Thus, focusing on the hypothetical single intervention on treatment $A$ does not suffice to identify causal effects in the target population. As noted in prior work [9], that would remain true even if the $S \rightarrow Y$ arrow were removed from the target population DAG in Figure 1.

Next, we might consider joint intervention to set trial participation $S$ to $s = 0$ and treatment $A$ to $a$, as found in Figure 3. The difficulty here is that, while the independence $Y^{s=0,a} \perp\!\!\!\perp S|X$ holds, neither $Y^{s=0,a} \perp\!\!\!\perp A^{s=0}|X$ nor $Y^{s=0,a} \perp\!\!\!\perp A^{s=0}|(X,S)$ hold because of the $A^{s=0} \leftarrow U \rightarrow Y^{s=0,a}$ path, again reflecting the presence of unmeasured confounding for treatment in the context of usual care implemented in the target population. Thus, in this causal structure the effect of treatment in the context of usual care is not identifiable, at least not without making additional assumptions.

Some progress is possible by considering the joint intervention to set trial participation $S$ to $s = 1$ and treatment $A$ to $a$ as found in Figure 4. Under this intervention, the $U \rightarrow A^{s=1}$ arrow is removed because under intervention to set trial participation $S$ to $s = 1$, treatment would not depend on unmeasured variables $U$. Thus, the effect of both components of the joint intervention of trial participation and treatment is identifiable by exploiting the independencies $Y^{s=1,a} \perp\!\!\!\perp A^{s=1}|(X,S)$ and $Y^{s=1,a} \perp\!\!\!\perp S|X$. That effects of joint interventions on trial participation and treatment are identifiable in the presence of trial engagement effects has also been noted in previous work [8]. Nevertheless, the effects of joint interventions to set trial participation $S$ to $s = 1$ and treatment $A$ to $a$ may not be practically relevant when the aspects of trial engagement that affect the outcome not through treatment are not available in the target population due to practical constraints, or because these aspects are not well-characterized.

Thus presents an identification conundrum: our scientific and clinical interest may lie in the identification of the average treatment regardless of the context of care, $\mathrm{E}[Y^{a=1} - Y^{a=0}]$, or the average treatment in the context of usual care as implemented in the target population, $\mathrm{E}[Y^{s=0,a=1} - Y^{s=0,a=0}]$, but these effects are not identifiable under the assumptions encoded in the DAG of Figure 1. Conversely, the causal estimand that captures the average treatment effect under trial protocol-based care, $E[Y^{s=1,a=1} - Y^{s=1,a=0}]$, is identifiable in Figure 4 but is arguably of lesser interest.

To address this identification problem, we will show in the next section that identification of $E[Y^{s=0,a=1} - Y^{s=0,a=0}]$, that is the average treatment effect in the context of usual care in the target population, is possible if background knowledge supports certain additional assumptions of no causal interaction between trial participation $S$ and treatment $A$ on the absolute and relative scales. The approaches we present also extend to estimands that pertain to the nonrandomized subset of the target population, such as the average treatment effect in the context of usual care among trial nonparticipants, $E[Y^{s=0,a=1} - Y^{s=0,a=0}|S=0]$; we present these results in the Supplement.

## 5 IDENTIFICATION WHEN THERE IS NO INTERACTION BETWEEN TRIAL PARTICIPATION AND TREATMENT ON THE ABSOLUTE SCALE

In this section, we provide identification results for the average treatment effect in the context of usual care as implemented in the target population. We show that identification is possible under progressively weaker sets of novel identifiability conditions that, because trial engagement effects are not assumed to be absent, pertain to joint interventions on trial participation $S$ *and* treatment $A$. Specifically, our first identification result relies on a condition of no causal interaction at the *individual-level*, between trial participation $S$ and treatment $A$ on the absolute scale, as well as conditional mean exchangeability (transportability) over trial participation $S$ (condition A2). Our second result relies on a condition of no interaction *in expectation*, between trial participation $S$ and treatment $A$ on the absolute scale (subsumed by the analogous individual-level condition in the first approach), as well as conditional mean exchangeability over trial participation $S$ (condition A2). Our third result relies on the same condition of no interaction in expectation between trial participation $S$ and treatment $A$ on the absolute scale, but uses an assumption of no effect measure modification by trial participation for the joint intervention on trial participation $S$ and treatment $A$, instead of conditional mean exchangeability over trial participation $S$ (i.e., substituting an assumption of no effect measure modification for condition A2).

Definitions of the terms "causal interaction" and "effect measure modification" as we use them in this paper can be found in the Supplement (Section 2). In the main text we only present results using nested expectation expressions (g-formula-like [29]), and provide weighting re-expressions in the Supplement (Sections 4.1.4 and 4.2.4). In all our identification results, the average treatment effect in the context of usual care as implemented in the target population is identified with the same observed data functionals as when trial engagement effects are assumed to be absent [1, 2, 4, 5, 35]. For this reason, we do not provide estimation or inference results because these have been addressed extensively by previous work.

## 5.1 Identification when there is no interaction between trial participation and treatment at the individual level

To fix ideas, we first consider the setting where there is no interaction between trial participation $S$ and treatment $A$ for each individual on the absolute scale. We begin by considering the following identifiability condition:

A6. *No individual-level interaction between trial participation and treatment on the absolute scale.* For every individual, there is no causal interaction between trial participation $S$ and treatment $A$ on the absolute scale, that is,

$$Y^{s=1,a=1} - Y^{s=1,a=0} - Y^{s=0,a=1} + Y^{s=0,a=0} = 0.$$

Condition A6 states that for every individual, the effect of intervening to set treatment $A$ to $a = 1$ versus $a = 0$ is the same, whether intervention on treatment occurs in the trial context (where $S$ is set to $s = 1$) or outside of it (where $S$ is set to $s = 0$).

The following proposition is our key identification result provided in the main text (proof available in Supplement Section 4.1.1):

**Proposition 1.** *Under conditions A1 through A6, the average treatment effect in the context of usual care in the target population, $\mathrm{E}[Y^{s=0,a=1} - Y^{s=0,a=0}]$, is identified with*

$$\mu = \mathrm{E}\Big[\mathrm{E}[Y|X, S = 1, A = 1] - \mathrm{E}[Y|X, S = 1, A = 0]\Big].$$

Here, the identifying functional $\mu$ is the same as in previous work on transportability methods [1, 4, 6, 35], where assumptions are made, either implicitly or explicitly, about the absence of trial engagement effects. In this setting, $\mu$ can be thought of as the conditional mean difference within the index trial standardized to the covariate distribution of the target population.

## 5.2 Identification when there is no interaction, in expectation, between trial participation and treatment

For settings where one may not be prepared to assume the absence of individual-level interactions, we now show that $\mu$ is identified with the same observed data functional under the condition of no interaction, in expectation, between trial participation $S$ and treatment $A$ on the absolute scale, conditional on covariates $X$. Specifically, we consider the following identifiability condition, which is weaker and is implied by condition A6 (though the reverse is not necessarily true):

A7. *No interaction, in expectation, between trial participation and treatment on the absolute scale.* In expectation, there is no causal interaction between trial participation $S$ and the randomized treatment $A$ on the absolute scale in the target population, conditional on covariates $X$, that is, for every covariate pattern $x$ with positive density $f(x) \neq 0$,

$$\mathrm{E}[Y^{s=1,a=1} - Y^{s=1,a=0}|X=x] = \mathrm{E}[Y^{s=0,a=1} - Y^{s=0,a=0}|X=x].$$

This condition states that, in expectation, the effect of intervening to set treatment $A$ to $a = 1$ versus $a = 0$ is the same, whether intervention on treatment occurs in the trial context (where $S$ is set to $s = 1$) or the context of usual care (where $S$ is set to $s = 0$), conditional on covariates $X$. This condition encodes a common approach to interpreting clinical trials, where the effect of randomized treatments is implicitly assumed to be separate to the effect of other trial-related procedures, conditional on covariates $X$.

Using condition A7, we obtain the following identification result (proof available in the Supplement Section 4.1.2):

**Proposition 2.** *Under conditions A1 through A5, and A7, the average treatment effect in the context of usual care in the target population, $\mathrm{E}[Y^{s=0,a=1} - Y^{s=0,a=0}]$, is also identified by $\mu$.*

To our knowledge, the identification results under the assumption of no interaction, in expectation, between trial participation and treatment have not been previously described in prior work. In applied work, this identification approach would allow for outcome-relevant trial procedures to differ from those implemented in the context of usual care in the target population, thus representing a meaningful relaxation of the assumption that trial engagement effects are absent.

## 5.3 Identification when there is no effect measure modification for the intervention comprising trial participation and treatment

Suppose again there is no interaction, in expectation, between trial participation and treatment on the absolute scale (condition A6). We show that the causal estimand of interest can identified even if we invoke a condition weaker than conditional exchangeability (transportability) over trial participation $S$ (condition A2). It suffices for there to be no effect measure modification by trial participation $S$, conditional on covariates $X$, for the joint intervention where trial participation $S$ is set to $s = 1$ and treatment $A$ is set to $a$. Specifically, we consider the following identifiability condition:

A8. *No effect modification for the joint intervention comprising trial participation and treatment on the absolute scale.* For the joint intervention to set trial participation $S$ to $s = 1$ and treatment $A$ to $a$, there is no effect measure modification by trial participation $S$ on the absolute scale, conditional on covariates $X$; that is, for every covariate pattern $x$ with positive density $f(x) \neq 0$,
$\mathrm{E}[Y^{s=1,a=1} - Y^{s=1,a=0}|X = x] = \mathrm{E}[Y^{s=1,a=1} - Y^{s=1,a=0}|X = x, S = 1]$.

Condition A8 states that the magnitude of the joint effect of scaling up trial participation activities and treatment would be the same in the index trial as in the target population (comprising both trial and non-trial participants), conditional on covariates $X$. The claim of no effect measure modification is weaker than the distributional assumption of conditional exchangeability over trial participation $S$ (condition A2), in the sense that the former is implied by the latter, but the reverse is not necessarily true. Furthermore, the plausibility of condition A8 may depend in part on whether the causal interaction variables for which trial participation $S$ may be a proxy feature among the covariates $X$ or not.

Previous work in generalizability and transportability methods has considered assumptions of no effect measure modification – either informally [36–38] or formally using potential outcomes [4, 5, 39] – though in settings where trial engagement effects were assumed to be absent. Because trial engagement effects were assumed to be absent, claims of no effect modification were made pertaining to intervention only on treatment, rather than on both trial participation and treatment as presented here. Nonetheless, condition A8 could be viewed as reflecting a common approach to the interpretation of trials and how their results should be applied to a target population of substantive interest. In some contexts, the trial and target populations might be sufficiently alike to justify the claim that, within levels of covariates, the effect of implementing outcome-relevant trial procedures and treatment does not vary between the trial and non-randomized populations.

Using condition A8 instead of A2, we obtain the following identification result (proof available in the Supplement Section 4.1.3):

**Proposition 3.** *Under conditions A1, A3 through A5, A7 and A8, the average treatment effect in the context of usual care in the target population, $\mathrm{E}[Y^{s=0,a=1} - Y^{s=0,a=0}]$, is also identified with $\mu$.*

Different from previous work, the identification results provided here demonstrate that under causal structures where trial engagement effects are present, identification cannot be reduced to a sole assessment of whether effect measure modification exists between the trial and target population, conditional on covariates. Instead, identification of the average treatment effect in the context of usual care as implemented in the target population requires a pair of assumptions: one about the absence of interaction between trial participation and treatment, *and* another about the absence of effect measure modification by trial participation on joint interventions on participation and treatment. More broadly, this result, as well as those presented above, support the notion that the problems of extending (generalizing or transporting) inferences from trials to a target population should not be viewed merely as standardization exercises, but instead as a new class of causal problems that require explicit consideration of the underlying causal structures.

# 6 IDENTIFICATION WHEN THERE IS NO INTERACTION BETWEEN TRIAL PARTICIPATION AND TREATMENT ON THE RELATIVE SCALE

In the Supplement (Section 5), we show that the identification results above can be extended to a setting where assumptions about the absence of causal interactions are made on the relative, instead of absolute, scale. To pursue these analyses, modifications are required to the sampling design and the data structure, including the need for target population data on outcomes under the control ($A = 0$) treatment. We also show, using basic algebraic arguments, that claims of no causal interaction on both the absolute and relative scales are unlikely to hold simultaneously unless other, stronger, assumptions hold; our arguments parallel prior work on effect measure modification [33] as well as well-known results about absolute and relative measures of association [40]. Therefore, investigators need to apply their background knowledge to determine the scale, if any, on which they are willing to assume the absence of causal interaction.

# 7 DISCUSSION

Much of the existing work on generalizability and transportability rests on implicit or explicit assumptions that trial engagement effects are absent. These assumptions allow for the effect of treatment to be identified without regard to the specific context in which treatment is administered, whether experimental or non-experimental [1, 2, 4, 5, 22, 41–43]. In many settings, however, it is commonly believed that the outcomes of trial participants may be affected by trial activities that are separate from the randomized treatments to the extent that trial participation could be thought of as an intervention separate from treatment assignment (or treatment). Here, we offer an alternative approach to identify a causal estimand of interest in the target population when trial engagement effects are present. Our main results show that the average treatment effect in the context of usual care as implemented in the target population can be identified under assumptions of no interaction between trial participation and treatment, offering a way forward for generalizability and transportability analyses even when trial engagement effects are thought to be present.

To build some intuition about our results, it may help to consider the following example. Suppose we wish to transport causal inferences from a randomized trial that compared a statin versus placebo on the 5-year risk of all cardiovascular events to a target population of treatment-eligible individuals with no history of coronary heart disease. Suppose further that standards of care with regard to blood pressure control differs between the randomized trial and the target population, a scenario often encountered in clinical practice. Using conventional methods, and as demonstrated in our discussion of the causal models presented, the presence of this trial engagement effect would render the average treatment effect non-identifiable. As noted in previous work [8, 9], without additional assumptions we would only be able to identify the joint effect of scaling up trial-associated activities (here, applying the standard of care for blood pressure management in the trial) and initiating statin therapy in the target population.

Using the results presented in this paper, progress is possible if we can assume the absence of causal interaction between trial participation and treatment. While this is still a strong assumption [44, 45], in some settings domain-specific knowledge could be considered to evaluate its plausibility. For instance, in our statin example, some relevant evidence is provided by the Heart Outcomes Prevention (HOPE)-3 trial, a large 2-by-2 factorial trial that randomized over 12,000 intermediate-risk study participants with no history of cardiovascular disease to (1) rosuvastatin 10mg daily versus (2) placebo, and to either (1) a combination of candesartan 16mg plus hydrochlorothiazide 12.5mg daily versus (2) placebo [46–48]. The trial found no strong evidence of statistical interaction between any of the randomized interventions on the relative scale. As usual, interpreting the results of empirical examinations of interaction requires considering the potential for limited precision, the scale on which interaction is assessed, and model specification [49–52]. Nevertheless, such assessments of interaction might contribute to subject matter deliberations about whether any outcome-relevant procedures that accompany trial participation causally interact, on average, with the primary treatment of interest, and whether one would be justified in pursuing a generalizability or transportability analysis in such a setting.

From a methodological perspective, the identification results herein also highlight the distinction between *causal interactions* between trial engagement and the randomized treatments, and *effect measure modification* (of the effect of treatment) by trial participation, two concepts rooted in epidemiologic tradition [16–18, 53–58]. We explicitly considered joint interventions on trial participation – entailing a set of actions that may have causal effects on the outcome not through treatment – and treatment, and showed that the claim of no interaction between trial participation and treatment could be coupled with the usual assumption of conditional exchangeability over trial participation to support identification. We then showed, as would be expected from prior work [4, 5, 37, 38, 59–61], that the latter condition could be relaxed substantially by claiming that the effect of the joint intervention did not vary between trial and non-trial participants, conditional on covariates. The claim of no effect measure modification for the joint intervention can be supported by study design, if trial participants are randomly chosen from the target population (e.g., in an experiment among randomly sampled individuals [62]). Alternatively, one may attempt to carefully measure all relevant effect measure modifiers in the index trial and adjust for these accordingly to target population (e.g., using standardization or weighting approaches). Invoking claims of no effect measure modification is closely tied to previously articulated approaches to identification that rely on transporting counterfactual risk differences or ratios from the index trial to the target population of interest [33, 39, 63, 64].

In summary, we have shown that generalizability and transportability analyses can identify average treatment effects in the context of usual care as implemented in the target population, even in the presence of trial engagement effects. These novel identification results rely on the condition of no interaction between trial participation and treatment on either the absolute or relative scales. Our findings suggest that estimators proposed in earlier work may have an alternative interpretation under the assumptions invoked in this paper. Beyond transporting the results of a single trial, our findings may have implications for studies that derive data from multiple sources, including multicohort studies (e.g., multicenter trials and causal meta-analyses [39]), where the potential for treatment engagement effects compound with the involvement of multiple study sites. While assumptions about causal interactions (and, where applicable, effect measure modification) are largely subjective in nature, careful consideration of the underlying causal structure of the problem may still allow average treatment effects to be identified and estimated in a target population of substantive interest.

# FIGURES

Figure 1: Causal directed acyclic graph (DAG) encoding relationships between baseline covariates $X$, trial participation $S$, treatment assignment $A$, non-failure time outcome $Y$, and unmeasured covariates $U$.

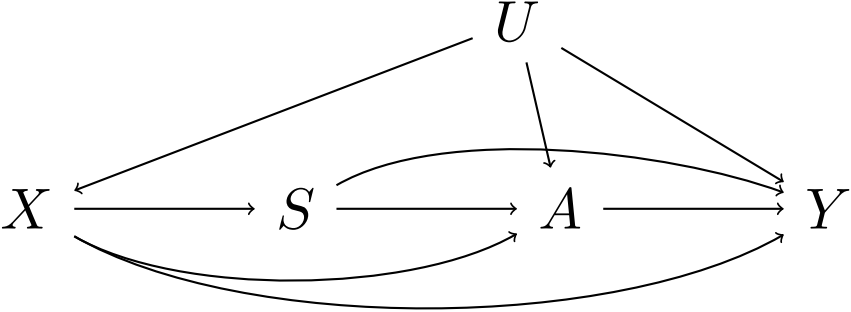

Figure 2: Single world intervention graph (SWIG) under intervention to set treatment assignment $A$ to $a$.

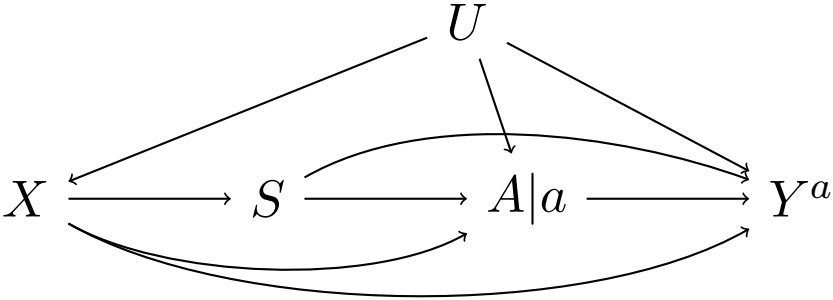

Figure 3: SWIG under joint intervention to set trial participation $S$ to $s = 0$ and treatment assignment $A$ to $a$.

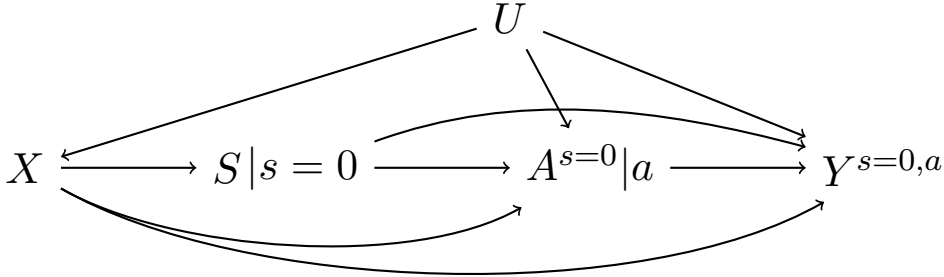

Figure 4: SWIG under joint intervention to set trial participation $S$ to $s = 1$ and treatment assignment $A$ to $a$.

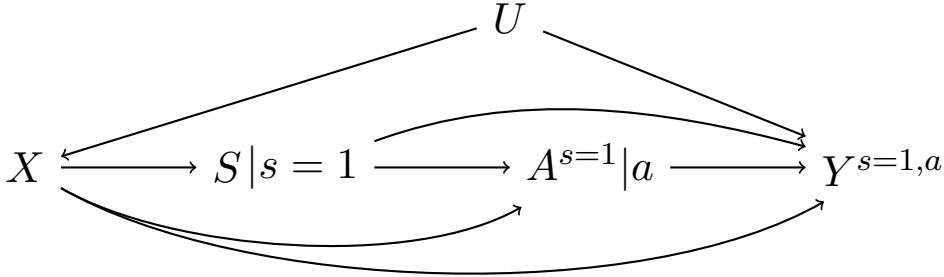